\documentclass{article}

\usepackage{arxiv}
\usepackage[utf8]{inputenc} 
\usepackage[T1]{fontenc}    
\usepackage{hyperref}       
\usepackage{url}            
\usepackage{booktabs}       
\usepackage{amsfonts}       
\usepackage{nicefrac}       
\usepackage{microtype}      
\usepackage{cleveref}       
\usepackage{graphicx}
\usepackage{doi}
\usepackage{enumerate}
\usepackage{xurl}
\usepackage{natbib}
\usepackage{placeins}
\usepackage{adjustbox}

\title{dcnnv-{\Large 19}: A deep convolutional neural network for covid-{\Large 19} detection in chest computed tomographies}

\date{}

\author{ \href{https://orcid.org/0000-0003-4541-2909}{\includegraphics[scale=0.06]{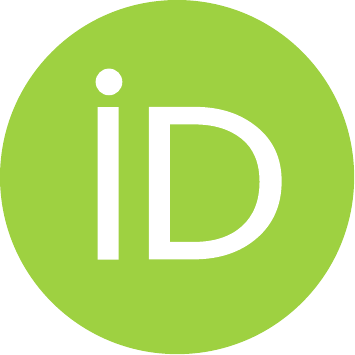}\hspace{1mm}Victor Felipe Reis-Silva} \\ GitHub: {\url{https://github.com/reis-victor/DCNNV-19}}\\São Paulo, São Paulo - Brasil \\ 2022
}

\hypersetup{
pdftitle={dcnnv-{\Large 19}: A deep convolutional neural network for COVID-{\Large 19} detection in chest computed tomographies},
pdfsubject={cs.CV, cs.AI, eess.IV},
pdfauthor={Victor Felipe Reis-Silva},
pdfkeywords={DCNN, COVID-19, SARS-CoV-2, tomography, convolutional},
}

\begin{document}
\maketitle

\begin{abstract}
This technical report proposes the use of a deep convolutional neural network as a preliminary diagnostic method in the analysis of chest computed tomography images from patients with symptoms of Severe Acute Respiratory Syndrome (SARS) and suspected COVID-19 disease, especially on occasions when the delay of the RT-PCR result and the absence of urgent care could result in serious temporary, long-term, or permanent health damage. The model was trained on 83,391 images, validated on 15,297, and tested on 22,185 figures, achieving an F1-Score of 98\%, 97.59\% in Cohen's Kappa, 98.4\% in Accuracy, and 5.09\% in Loss. Attesting a highly accurate automated classification and providing results in less time than the current gold-standard exam, \textit{Real-Time reverse-transcriptase Polymerase Chain Reaction} (RT-PCR).
\end{abstract}


\newpage
\tableofcontents
\newpage

\section{Introduction}
\subsection{The COVID-19 pandemic}

\subsubsection{Detection methods}

The gold standard test for COVID-19 diagnosis is the \textit{Real-Time reverse-transcriptase Polymerase Chain Reaction} (RT-PCR). \citep{tsang2021diagnostic} points out in their meta-study about sample types used in RT-PCR, that nasopharyngeal or combined nasal and throat swabs achieved the highest sensitivity, of 97\%, against saliva (85\%), nasal swabs (86\%), and throat swabs (68\%). The test could be performed on the first day of symptoms of a SARS, however, it is recommended to wait at least three days, due to the possibility of a low virus replication rate that could lead to a false negative. According to \citep{jawerth2020covid}, the test processing can be completed in up to three hours, but it is usually processed in a period of six to eight hours by most laboratories. Because of the high demand, the average time to deliver the test result is at least a business day.

Due to the RT-PCR waiting time, X-ray images and computed tomography (CT) have been used for pulmonary analysis in suspected cases of COVID-19 with moderate and severe respiratory symptoms, in order to act quickly in cases with possible evolution to hospitalization. 

Diagnostic imaging methods are also used as a complementary method to observe the evolution of the clinical picture of patients with lung involvement. CT scans provide better resolution and accuracy than X-ray images, producing three-dimensional images without overlapping of internal organs, even in small or thin structures. According to the review by \citep{karam2021chest} involving thirteen non-randomized studies of 4,092 patients, chest CT scan showed the following medians in sensitivity, specificity, and accuracy, respectively: 91\% (range 82\%-98\%), 77.5\% (range 25\%-100\%), and 87\% (range 68\%-99\%), with RT-PCR as the reference.
Considering these results, it is noted that despite the accuracy of 87\%, a higher rate of false positives (100\% - specificity = 22.5\%) than false negatives (100\% - sensitivity = 9\%) is observed, which in a medical context can be considered less serious, since in a case of false negative, a RT-PCR could be requested by the responsible physician in case of suspicion and the patient could be kept under observation until the diagnostic confirmation. 

\subsubsection{Chest CT findings in COVID-19 patients}

According to \citep{rosa2020covid}, the characteristics usually found in CT scans of patients with pneumonia caused by COVID-19 are:

\begin{enumerate}[a.]
\item Ground-glass opacity: the most commonly found pulmonary alteration, visible from the onset of respiratory symptoms until approximately the fourth day. It occurs due to increased lung density without bronchovascular degradation and usually occurs in the lower lobes in a bilateral peripheral and subpleural configuration; 

\item Crazy paving: characterized by the superimposition of ground-glass opacity by the thickening of the interlobular and intralobular septa; 

\item Pleural effusion: observed in severe cases, it is characterized by excessive accumulation of fluid in the pleural region; 

\item  Reversed halo sign: found when a ring of consolidation surrounds a rounded region of ground-glass opacity; 

\item Consolidation: occurs when the alveoli are filled with inflammatory exudate, making the region opaque on imaging exams; 

\item Air-bronchogram: visualized when the air-filled bronchi are visible due to consolidation of the adjacent alveolar region. 
\end{enumerate}
 
Examples of each of these cases can be seen in Figure 1: 

\begin{figure}[htbp]
\makebox[\linewidth]{\includegraphics[width=0.85\linewidth]{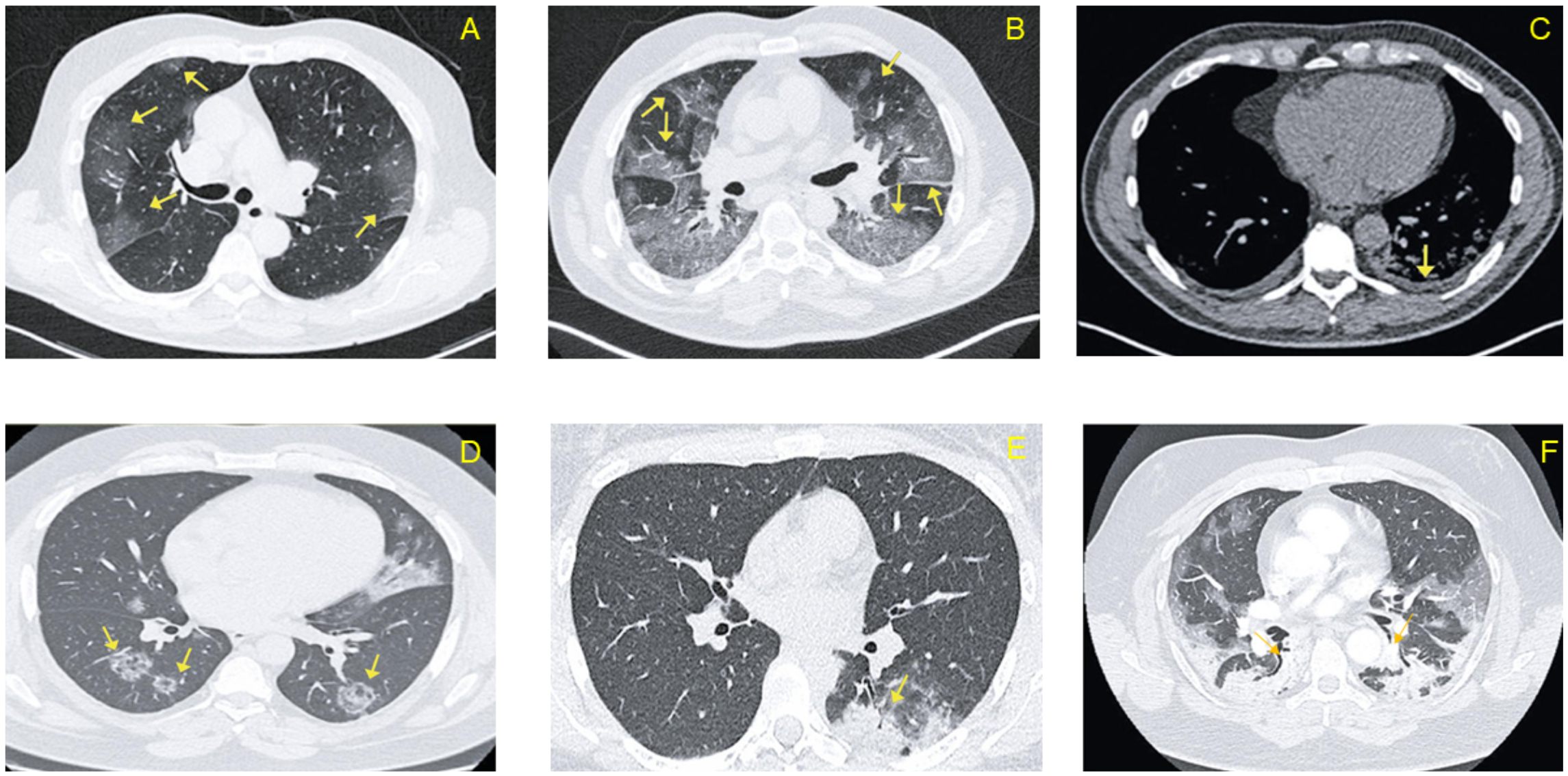}}
\caption{Chest CT findings in COVID-19 patients. From left to the right: Ground-glass opacity (A), Crazy paving (B), Pleural effusion (C),  Reversed halo sign (D), Consolidation (E), and Air-bronchogram (F). Source: Author compilation adapted from \citep{rosa2020covid}}
\end{figure}
\FloatBarrier

\subsection{Context in Machine Learning}

The use of Computer Vision techniques increased in recent years, mainly due to the advance in Deep Learning techniques and the rapid evolution and price decrease of hardware, especially the Graphics Processing Unit (GPU), popularly known as "graphics card", which have become more powerful and cheaper, and is the main component\footnote{In 2016, Google introduced the Tensor Processing Unit (TPU), a circuit designed specifically for processing Machine Learning tasks, optimizing the execution of these tasks when performed on the TensorFlow framework. The use of TPU has increased recently, mainly due to its advantage in processing speed.} used for processing tasks using artificial intelligence techniques.

In 1998, the first modern implementation of a Convolutional Neural Network, LeNet-5, by \citep{lecun1998gradient}, revolutionized the field of computer vision. This class of neural network was inspired by the Neocognitron, a multilayer artificial neural network published in 1982 by \citep{fukushima1982neocognitron}, based on the studies on the functioning of the animal visual nervous system proposed by \citep{hubel1959receptive} after studies with cats and monkeys, respectively in the periods 1959-1962 and 1968. The LeNet-5 has been widely used in the recognition of handwritten numbers in digital images, for example, in the recognition of numbers on bank checks and also in the recognition of ZIP codes in letters. Since its success, this class of artificial neural network specialized in processing data such as images and videos has been used increasingly for the automation of visual tasks. Medicine is one of the areas of increasing use due to the constant improvement of machine learning algorithms, which often match or surpass humans in accuracy, with the added benefit of performing tasks much faster and in an automated way, for example: detection, segmentation, and classification of cancer, skin diseases, tumors, bone fractures; visual monitoring of diseases, helping physicians to observe signs of regression or progression; visual aid in delicate microsurgery and real-time monitoring of blood loss during and after surgery; and even cell counting in clinical analysis. The choice of this class of neural network as a solution to the proposed problem is due to its ability to learn the most important features of an image in an unsupervised manner, for example, learning to identify a cat in an image based on features such as the shape of its eyes, whiskers, and ears. Another advantage is having the neurons only locally connected and shared parameters, resulting in much fewer parameters than a fully-connected neural network, requiring less computational cost, and unlike the latter, being able to detect a learned feature anywhere in an image.

\section{Data} 
\subsection{Characteristics}

The dataset used for the model is based on the COVIDx CT-2A, compiled by \citep{Gunraj2021} from public datasets of chest CT images of healthy patients, patients with classical pneumonia, or COVID-19. The dataset is licensed under Attribution-NonCommercial-ShareAlike 4.0 International (CC BY-NC-SA 4.0). 2,282 images from 751 patients with no confirmed diagnosis were removed, resulting in a dataset containing 192,640 PNG images of 2,994 patients, with a resolution of 512 x 512 pixels.

\subsection{Exploratory Data Analysis}
\subsubsection{Classes x Dataset}
The dataset is remarkably unbalanced, biased to the COVID-19 class (61.39\% of the total), being approximately four times larger than the Normal class and around two and a half times larger than the Pneumonia class. And in its original distribution, such bias also occurs in the selection of images for training, counterbalanced by the Normal class in the selection for validation and testing. The percentage of each class in the distribution can be seen in Figure 2:

\begin{figure}[htb]
\makebox[\linewidth]{\includegraphics[width=0.8\linewidth]{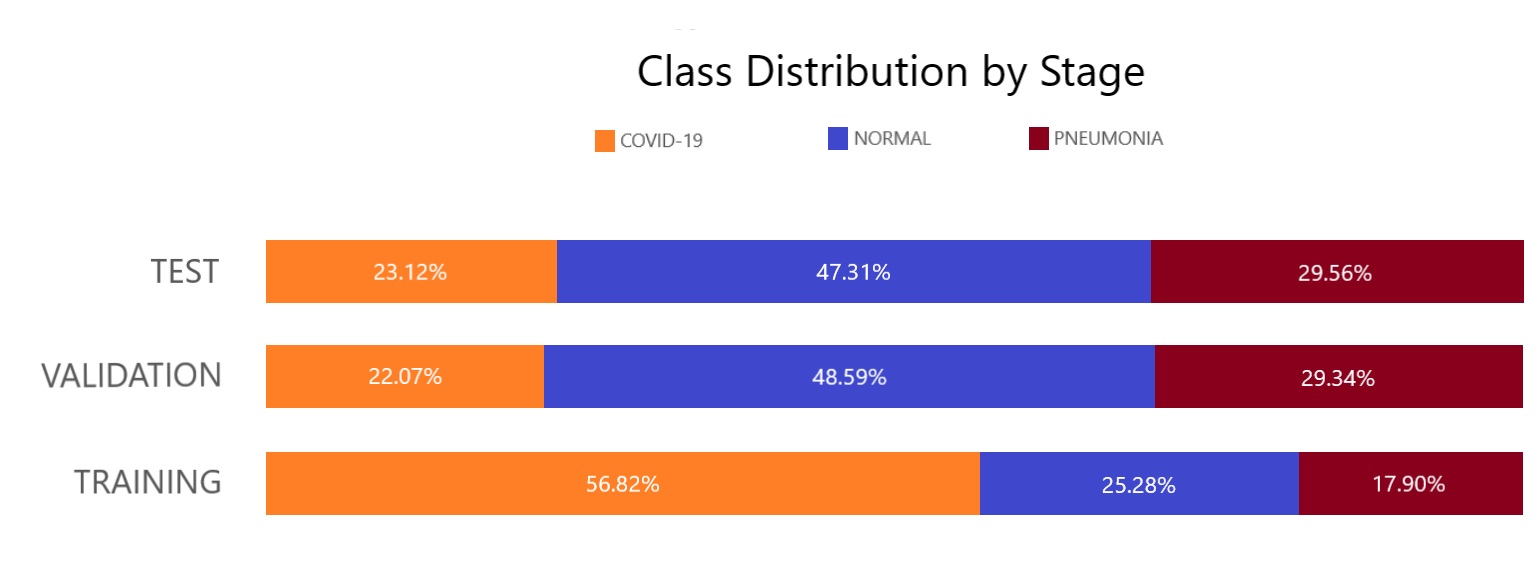}}
\caption{Class Distribution by Stage. Source: Author, adapted from the \citep{Gunraj2021} COVIDx CT-2 dataset}
\end{figure}
\FloatBarrier

\subsubsection{Classes x Country}

Looking at Table 1, it can be noticed that approximately 61\% of the images are of patients affected by COVID-19, and most of them are Chinese. Almost the entire sample (around 95\%) of people diagnosed with Pneumonia (23.31\% of the dataset) is also of Chinese nationality. In contrast, the samples of healthy individuals (15.30\%) is mainly divided between Iranian (49.21\%) and Chinese (42.41\%) people. Another point to consider is that the number of patients with no nationality informed (19.79\%) is greater than the sum of all nationalities except the Chinese.

\begin{table}[htb]
\makebox[\linewidth]{\includegraphics[width=0.8\linewidth]{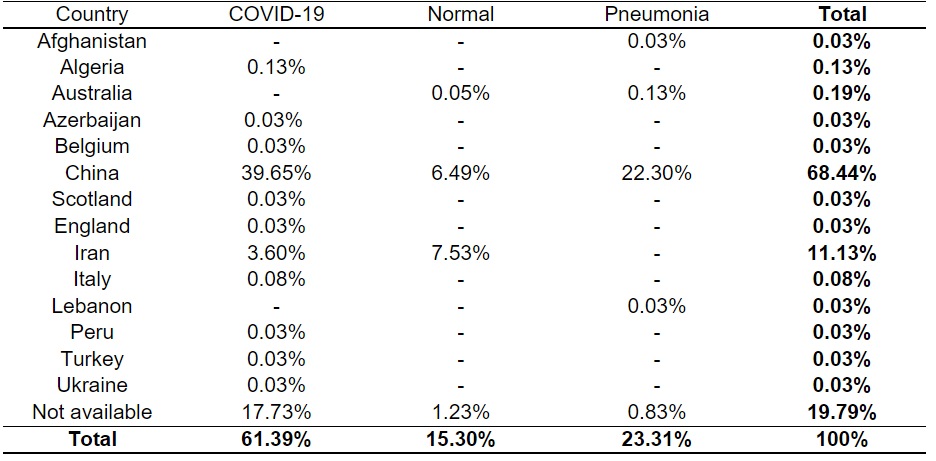}}
\caption{Class Distribution by Country. Source: Author, adapted from the \citep{Gunraj2021} COVIDx CT-2 dataset}
\end{table}
\FloatBarrier

\subsubsection{Diagnostics x Sex}

The information about the sex of the patient was available in only 39.35\% of the cases (19.57\% female and 19.79\% male), being absent for more than half (60.64\%) of the people, as can be seen in Figure 3:

\begin{figure}[htb]
\makebox[\linewidth]{\includegraphics[width=0.4\linewidth]{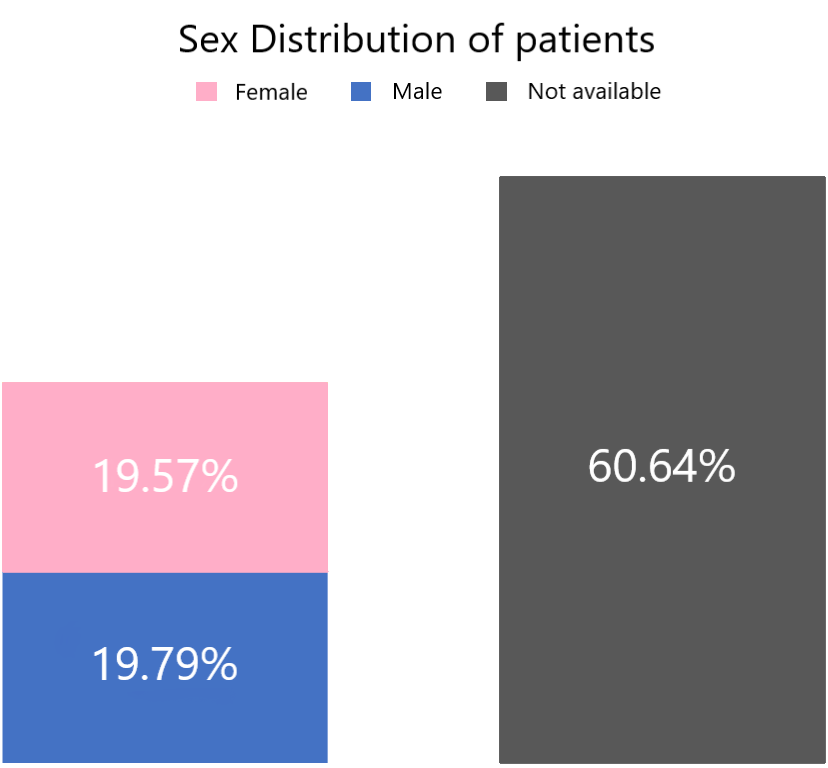}}
\caption{Sex distribution of the patients. Source: Author, adapted from the \citep{Gunraj2021} COVIDx CT-2 dataset}
\end{figure}
\FloatBarrier

\subsubsection{Diagnostics x Age}

Approximately 61\% of the individuals did not have their age informed. Among the 39\% of the individuals that had the age reported, it can be seen in Figure 4 that the age range between 0 and 20 years and the elderly over 81 years are a very small portion (both totaling only 2.4\% of the total), with a predominance of cases in the age range between 41 and 60 years:

\begin{figure}[htb]
\makebox[\linewidth]{\includegraphics[width=0.75\linewidth]{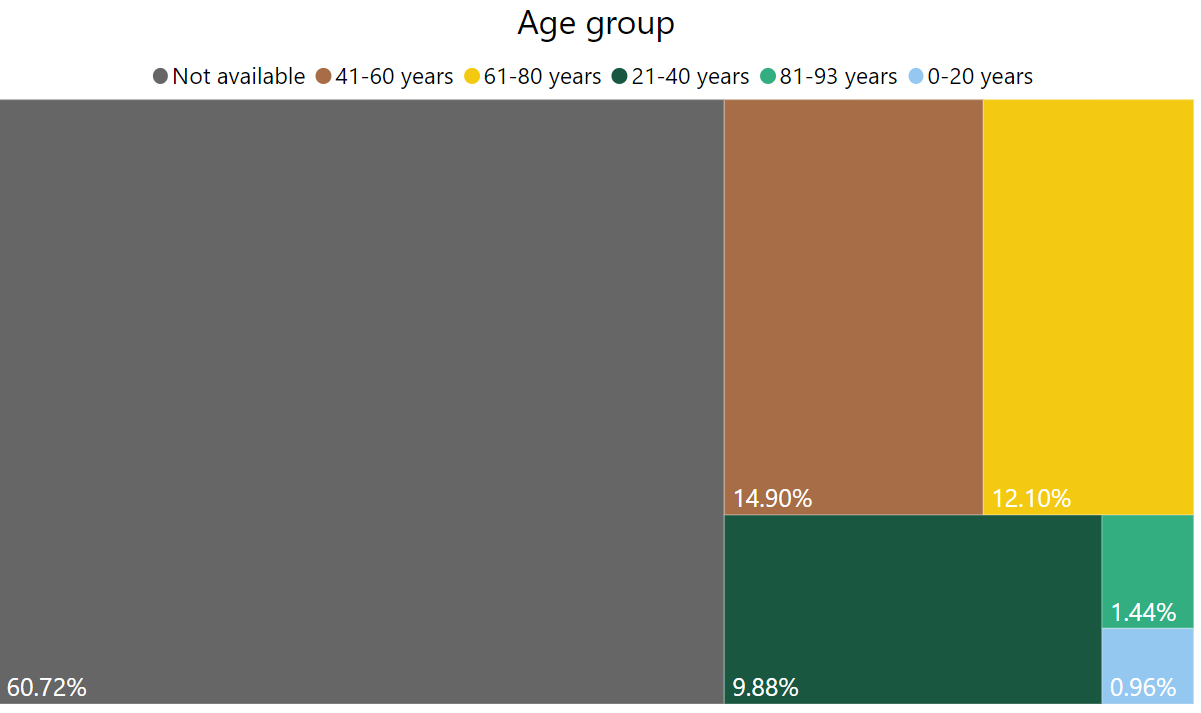}}
\caption{Class Distribution by age group. Source: Author, adapted from the \citep{Gunraj2021} COVIDx CT-2 dataset}
\end{figure}
\FloatBarrier

\section{Data treatment and Processing}

Image cropping was performed based on the information in the bounding boxes of each tomography, indicated in the files corresponding to each training stage. The adjustment aims to preserve only the essential portion of the image (removal of external structures to the area comprising the lung region). After the cropping, the images were resized from 512x512 pixels to 224x224 pixels and the resulting images were moved, following the metadata.csv file, to their respective directory (training, validation or test) and subdirectory with the respective number of their class (0, 1 or 2), and later the class folder names were changed to Normal, Pneumonia and COVID-19 respectively. At the end of the processing, the dataset size was reduced from 31.2 GB to 5.9 GB. The script can be found on project's GitHub. Figure 5 shows an example of an image after it has been cropped and resized: 

\begin{figure}[htb]
\makebox[\linewidth]{\includegraphics[width=0.6\linewidth]{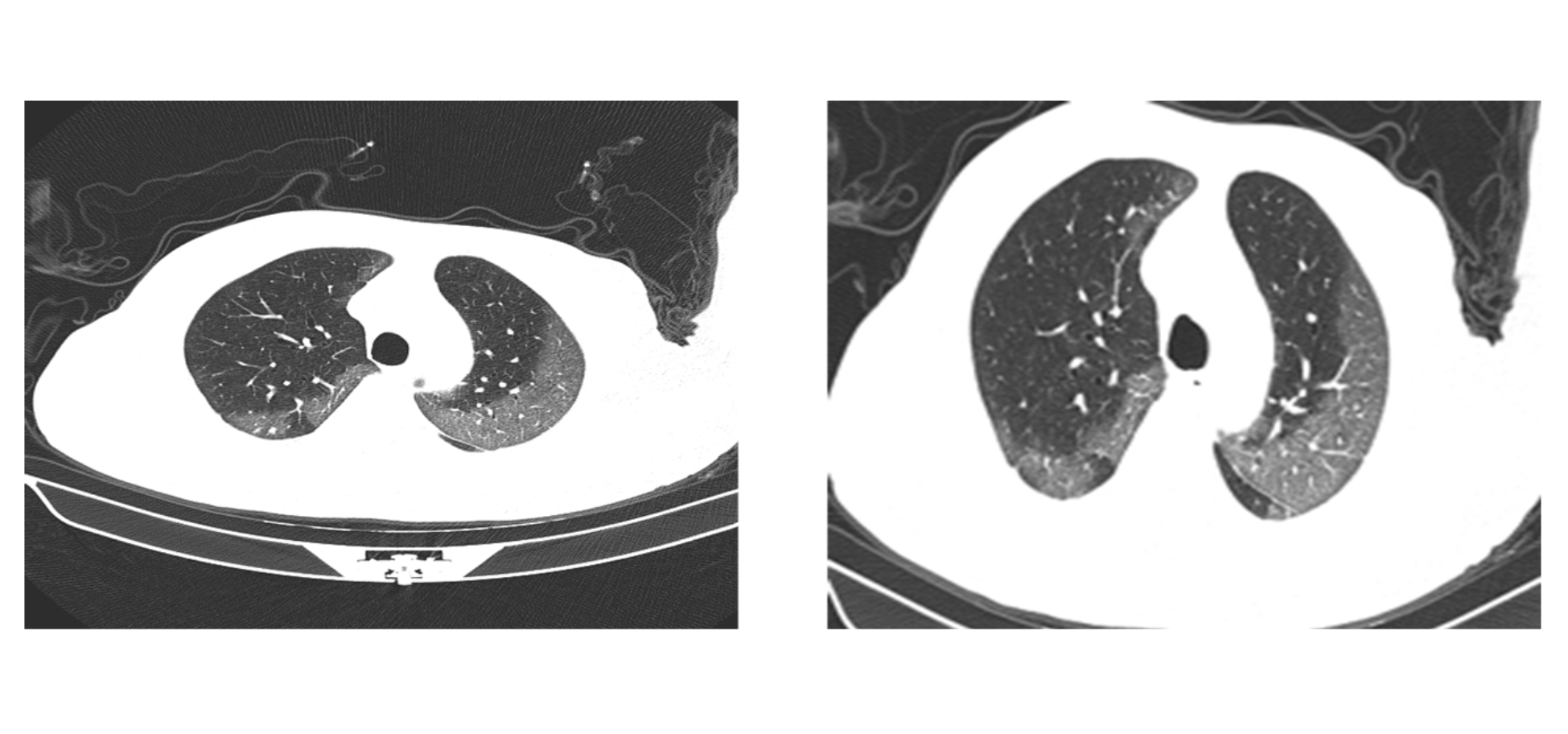}}
\caption{Comparison between an original image and its processed result}
\end{figure}
\FloatBarrier

\section{How do Convolutional Neural Networks work?}

\subsection{Architecture}    

Usually, a CNN is composed of 4 types of layers: convolutional layer, pooling layer, flattening layer and fully-connected layer. An example of the architecture of a convolutional neural network can be seen in Figure 6:

\begin{figure}[htb]
\makebox[\linewidth]{\includegraphics[width=1.0\linewidth]{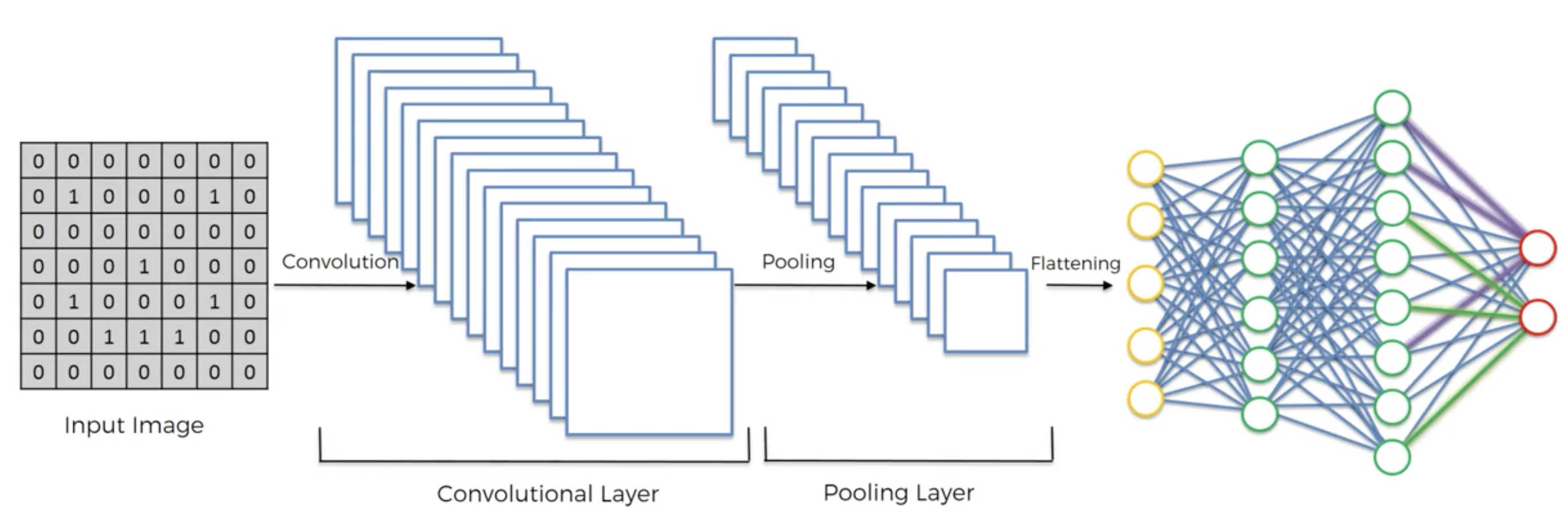}}
\caption{Example of a CNN architecture. Source: \citep{Mishra2022}}
\end{figure}
\FloatBarrier

\subsection{Structure and operation of a convolutional layer}

\subsubsection{Filters or kernels}
After receiving the image tensors in the input layer, the CNN tries to detect features by applying kernels, or filters. When dealing with a 2D filter, the use of these terms
is interchangeable, however, a 3D filter is composed of stacked kernels. Examples of different kernels in the recognition of the character "X" can be seen in Figure 7:

\begin{figure}[!htb]
\center
\makebox[\linewidth]{\includegraphics[width=0.45\linewidth]{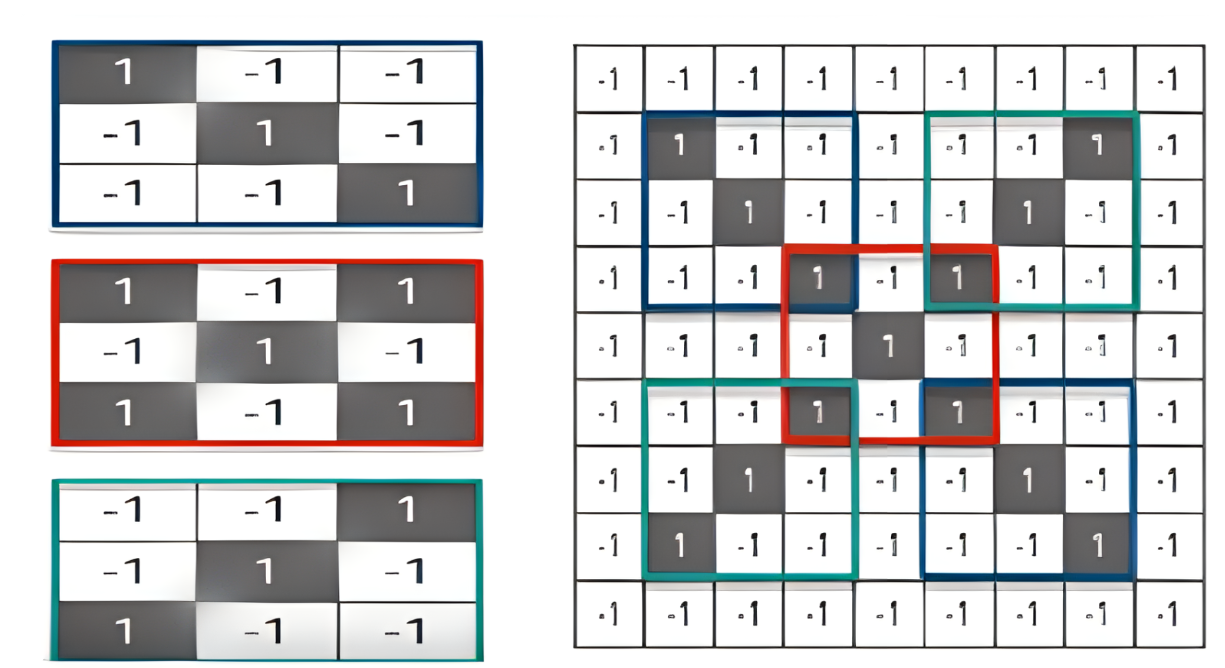}}
\caption{Examples of kernels in the recognition of the character "X". Each of the shaded rectangles (blue, red, and green) in the left column is a kernel responsible for detecting a feature. Source: adapted from \citep{Somavanshi2022}}
\end{figure}
\FloatBarrier

\subsubsection{Convolution}
The operation done by the filters to the input is based on the idea of convolution. However, the operation performed by most frameworks is a cross-correlation, due to its better performance by do not rotating the kernel 180 degrees with respect to the axes, characteristic of a convolution. The comparison between the mentioned operations can be seen in Figure 8:

\begin{figure}[htb]
\makebox[\linewidth]{\includegraphics[width=0.6\linewidth]{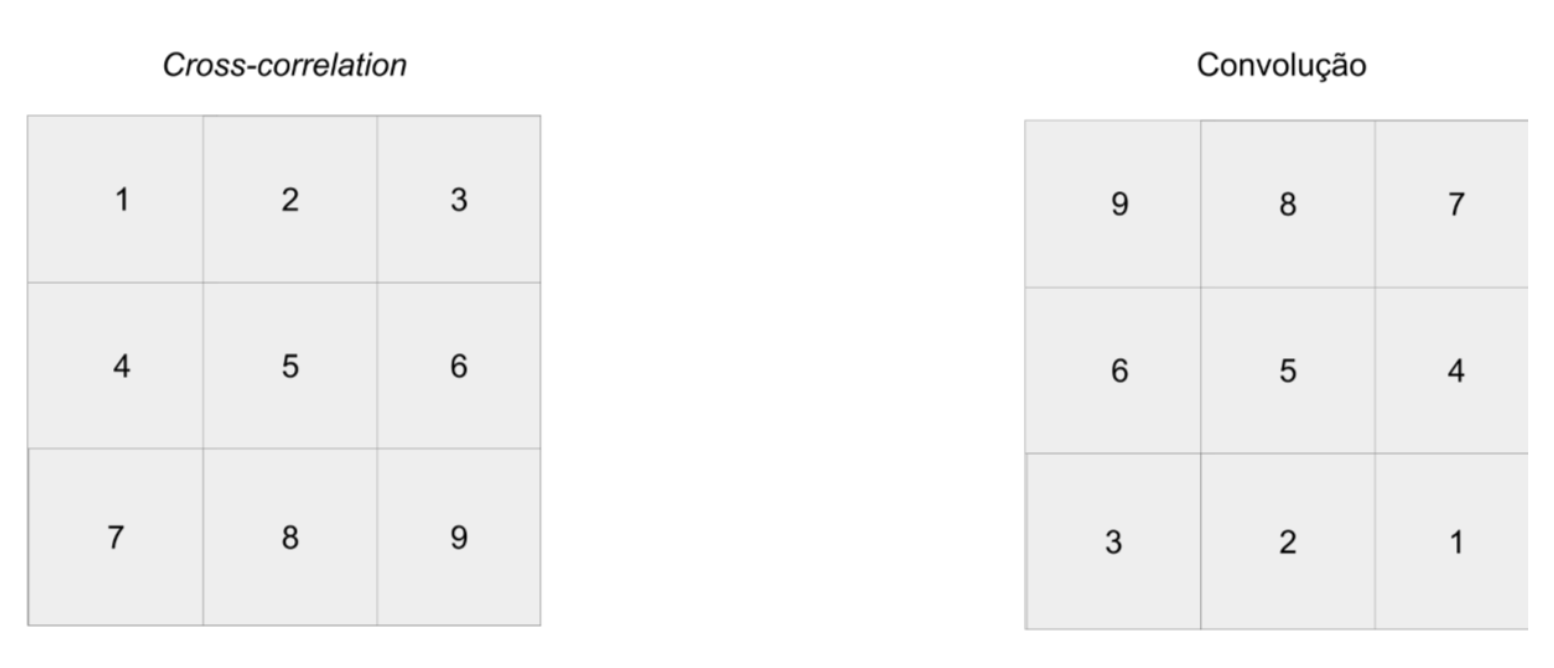}}
\caption{Comparison between Cross-correlation and Convolution. Source: Author [2022] }
\end{figure}
\FloatBarrier

During the convolution process, each kernel is aligned to a region of the input image of the same size, called "receptive field", and it is performed a multiplication of the values within this field, as can be seen in the example in Figure 9, which uses the blue-shaded kernel from Figure 9 in this process: 

\begin{figure}[htb]
\makebox[\linewidth]{\includegraphics[width=0.55\linewidth]{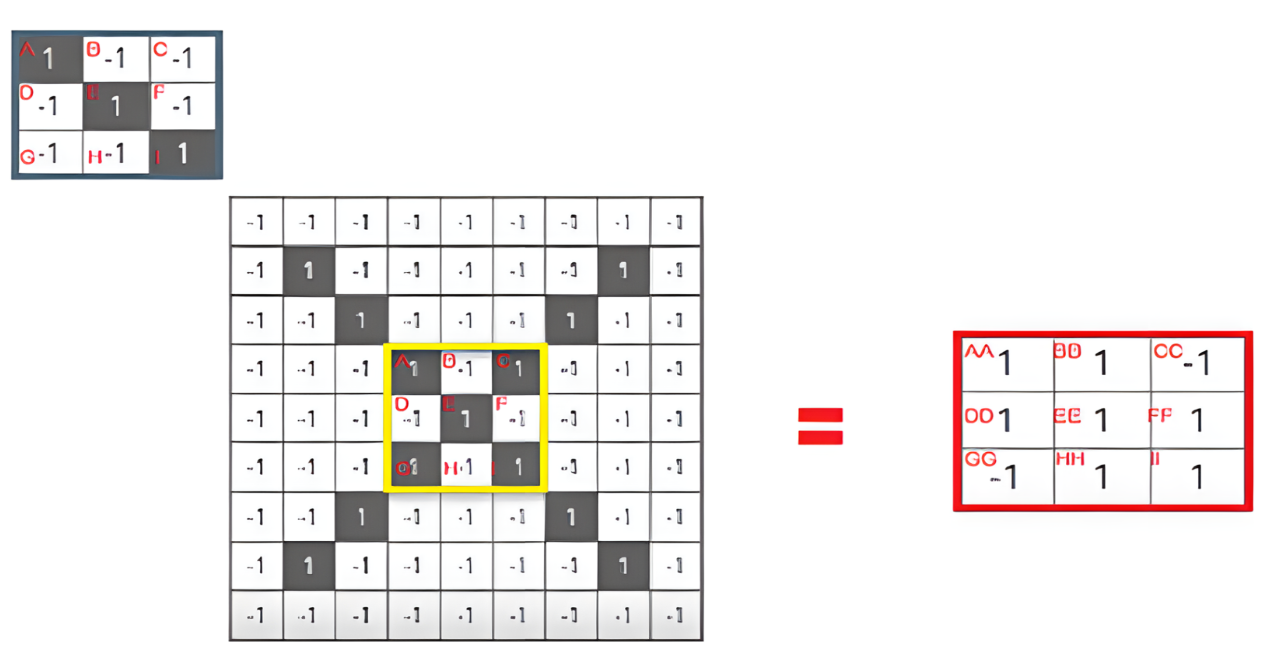}}
\caption{The 3 x 3 kernel (blue shaded) has its values multiplied by the corresponding locations of the 3 x 3 receptive field of the input image (yellow shaded square). Source: adapted from \citep{Somavanshi2022}}
\end{figure}
\FloatBarrier

At the end of the element-wise multiplication (A multiplied by A, B by B, and so on) the scalar product operation is performed, then the result of the multiplication of each of the points is added, and divided by the number of pixels. This process is shown in Figure 10:

\begin{figure}[htb]
\makebox[\linewidth]{\includegraphics[width=0.25\linewidth]{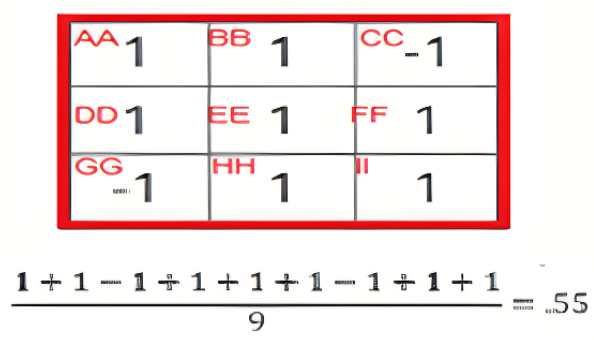}}
\caption{Convolution result. Source: adapted from \citep{Somavanshi2022}}
\end{figure}
\FloatBarrier

At the end of the convolution, the resulting value of the convolution operation of the receptive field is counted, Figure 11 indicates the location of the dot product on the feature map after the previously performed convolution operation: 

\begin{figure}[htb]
\makebox[\linewidth]{\includegraphics[width=0.65\linewidth]{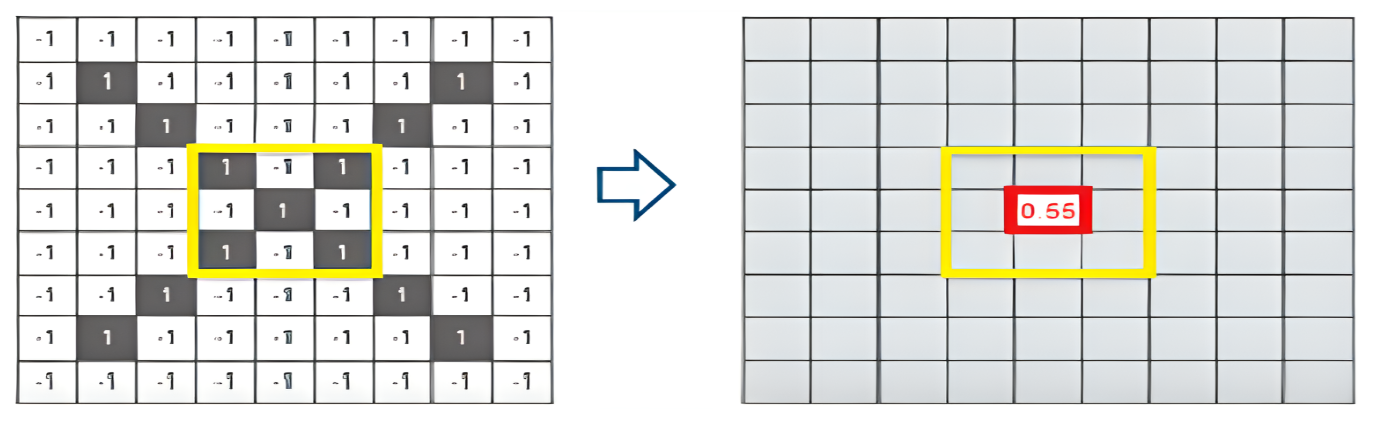}}
\caption{Dot product location at the end of the convolution process. Source: adapted from \citep{Somavanshi2022} }
\end{figure}
\FloatBarrier

The process is repeated until the filter has "slid" over the whole image. The way the sliding occurs depends on the values of two variables: 

\begin{itemize}

\item Stride: indicates how many units the filter should "slide" as it moves through the receptive field matrix.

\item Padding: process in which the input image value matrix is surrounded by a constant, usually zero, which is intended to prevent the decrease of matrix dimensions in the output (dimensionality reduction) and also to preserve/contribute to the learning of features present on the edges of the original image. 
\end{itemize}

At the end of the convolution process, the feature map or activation map is obtained. The activation map of the character X from the used blue shaded kernel can be seen in Figure 12:

\begin{figure}[htb]
\makebox[\linewidth]{\includegraphics[width=0.3\linewidth]{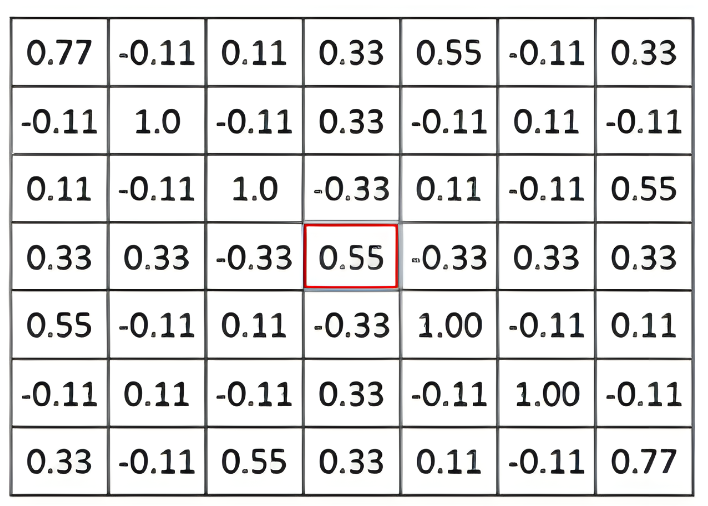}}
\caption{Feature map of the character "x" from the blue shaded kernel. Source: adapted from \citep{Somavanshi2022} }
\end{figure}
\FloatBarrier

\subsection{Activation function}
This function provides non-linearity to the convolutional layer, and its purpose is to perform the mathematical calculations needed to decide whether a neuron will be activated, thus it acts to control the output of the convolutional layer, and when the neuron is activated its information is passed to the subsequent layer.

\subsection{Pooling layer}
Usually, a convolutional layer is followed by a pooling layer, which receives it as input. In the pooling layer a subsampling process is performed, in which the spatial resolution of the feature map is reduced, decreasing the number of parameters and computational demands, and also contributing to the prevention of overfitting (when the model overfits the training data resulting in low generalization ability, being unable to generate good classifications using other datasets). The two most common clustering processes are average pooling and max pooling. The former averages the fields in the receptive field, unlike the latter, in which the highest value found in the receptive field is selected. Max pooling is the most used method, because it extracts the main features while keeping them rotation and position invariant. The calculation of both types is exemplified in Figure 13:

\begin{figure}[htb]
\makebox[\linewidth]{\includegraphics[width=0.55\linewidth]{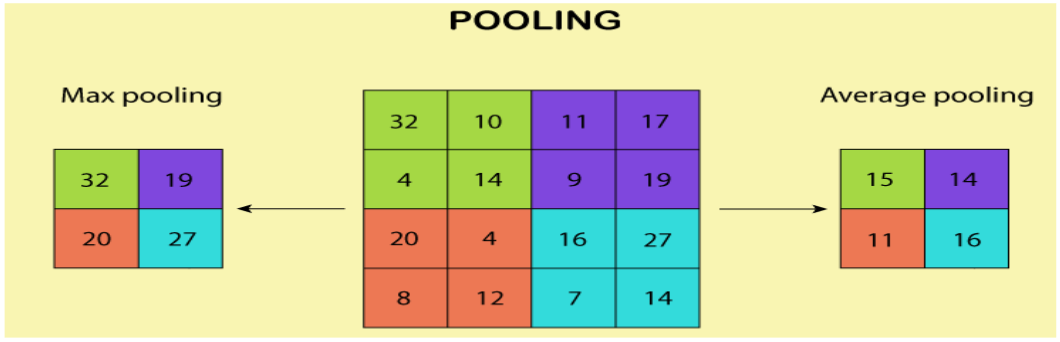}}
\caption{Max pooling and average pooling examples. Source: \citep{Swapna2020}}
\end{figure}
\FloatBarrier

\subsection{Flattening layer}
After the feature extraction step, the feature map matrix resulting from the last pooling layer is received by the flattening layer, which generates an output in the format of a (one-dimensional) vector. 

\subsection{Fully-connected layer or Dense layer}
This type of layer is fully connected to the neurons of the layer that comes before and after it. In a CNN, it receives the vector output coming from the flattening layer. At the end of a convolutional neural network, one or more fully-connected layers can be found, the last of which is called the output layer, which performs the classification.  

\section{Analysis}
\subsection{Model}
Figure 14 shows the architecture of the deep convolutional neural network model used in the dataset analysis, the final model files generated by TensorFlow are available on the project's GitHub. The model has 159,643 total parameters, of which 158,891 are trainable, and 752 are non-trainable.

\begin{figure}[htb]
\vspace{-1.1em}
\makebox[\linewidth]{\includegraphics[width=0.92\linewidth]{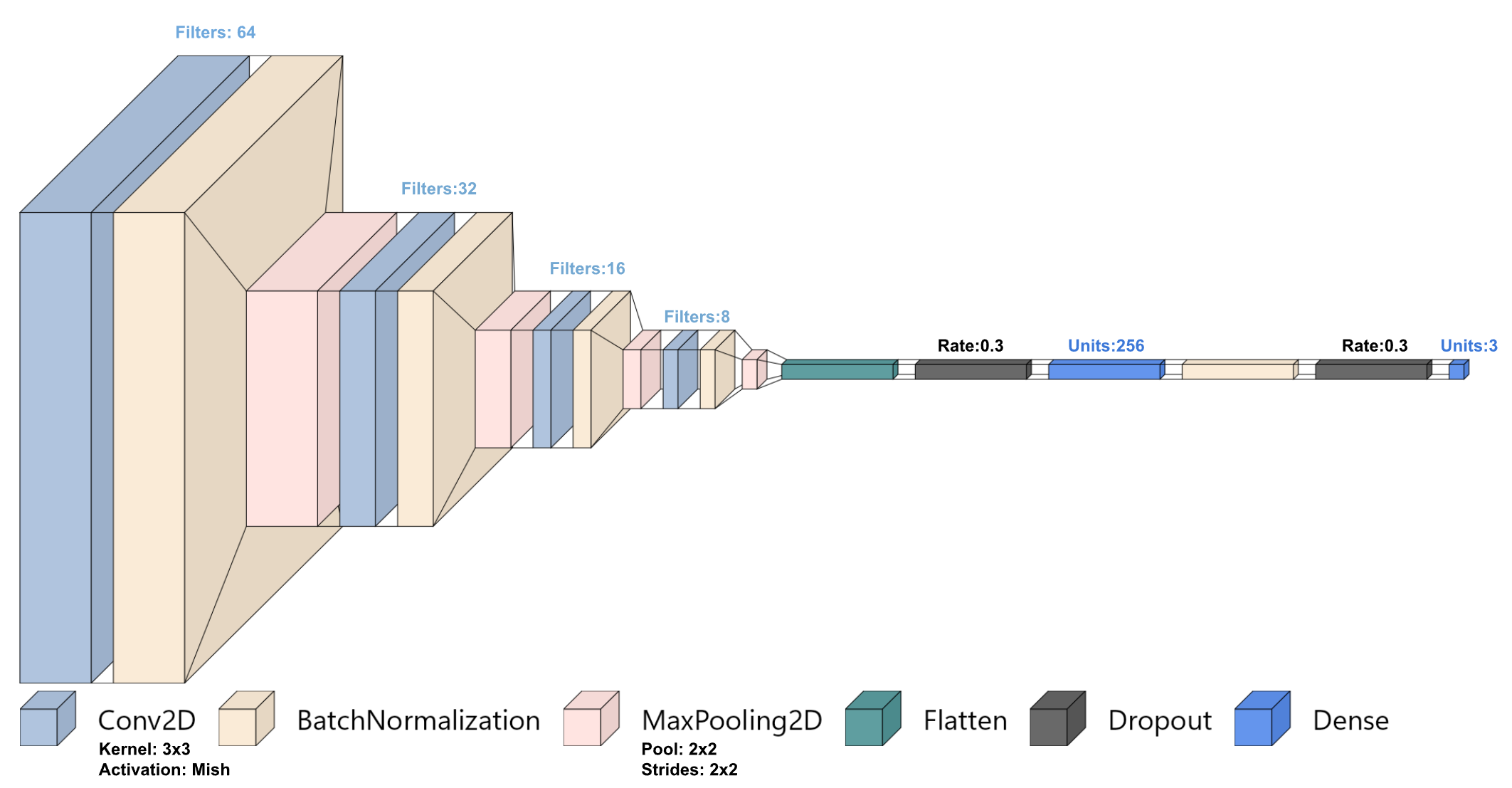}}
\caption{DCNNV-19 deep convolutional neural network model architecture. Source: Author, image generated with the package \textit{visualkeras}, by \cite{Gavrikov2020VisualKeras}}
\end{figure}
\FloatBarrier

\subsection{Standardization, Normalization, and Regularization}

An initial rescaling layer (\textit{tf.keras.layers.Rescaling}) was used to standardize the input values, rescaling the pixel values of the RGB channels (red, green, and blue) from the range 0-255 to 0-1, transforming the data to a smaller scale, making the model less computationally intensive and resulting in faster convergence of the neural network.						 
To improve the training stability and speed, Batch Normalization layers were applied after all convolutional layers and after the first dense layer. They can improve the neural network because of their capability of stabilizing aspects such as initialization and hyperparameters, resulting in a more independent learning of each layer. It also acts as a regularization layer, preventing overfitting. 

The Dropout regularization technique, which randomly turns off (turns the value to 0) neurons, was used to improve the generalization and to prevent overfitting that could occur due to the interdependence of features along the neural net.

\subsection{Parameters and Hyperparameters}

The images previously cropped and resized to 224 x 224 pixels were resized to 128 x 128 pixels, to enable the use of large batches, a computationally demanding task at full resolution for the GPU VRAM (Video Random Access Memory). The data augmentation processing (a process that creates new data/images applying changes to the data set such as: rotation, zooming, cropping, brightness changes, saturation, contrast, and coloring) showed no relevant advantage, besides increasing the training time.

To create a TensorFlow dataset, the function \textit{tf.keras.utils.image\_dataset\_from\_directory} was used, due to its better performance than \textit{ImageDataGenerator}, according to \citep{Gobeil2022}. The shuffle was kept as enabled, and its seed had a value manually chosen to conserve reproducibility. Converting the images for grayscale impacted accuracy and loss scores, thus the three color channels were preserved, even though the computational cost was higher.

Buffer prefetching was used(via the function \textit{Dataset.cache}, \textit{Dataset.prefetch}, and \textit{Dataset.AUTOTUNE}for dynamic tuning during execution) to avoid overloading and consequent slowdown due to disk input/output flow when performing operations with the dataset. 

Since the complete dataset has almost 200,000 images and is quite unbalanced, a selection was used, and the number of images in each class were equalized to the smallest class, resulting in the following configuration for each stage: training: 27,797 images per class, total of 83,391 images; validation, 5,099 images per class, total of 15,297; and testing, 7,395 images per class, a total of 22,185.	

The following optimizers were tested: \textit{Adam}, \textit{Adagrad}, \textit{Nadam}, \textit{SGD}, \textit{AdaBelief}, \textit{LAMB}, \textit{LazyAdam}, \textit{Rectified Adam}, \textit{Yogi}, and \textit{AdaBelief}. Being the \textit{AdaBelief} the optimizer with the best performance. As per the paper of its creators, \citep{DBLP:journals/corr/abs-2010-07468}, this optimizer has the great advantage of having the fast convergence achieved by adaptive methods such as \textit{Adam}, and also the high generalization capacity as does \textit{SGD}.	
Firstly, the Triangular Cyclical Learning Rate optimizer was used as the
learning rate parameter, to help to avoid problems caused by a learning rate that could be too low (which could cause learning not to progress), or an initial learning rate that could hinder the overall performance of model training. This optimizer varies the learning rate cyclically during steps (each iteration of the model with the selected batch size), its results can be used to find the optimal range between the lowest and highest learning rate by comparing the graphs of learning rate, accuracy and loss through the TensorFlow Tensorboard. After finding the best learning rate range, the maximum value was used as a parameter, in order to accelerate the learning, and the minimum value (in this case, the value suggested in the article \citep{DBLP:journals/corr/abs-2010-07468}, from the \textit{AdaBelief}) optimizer was added to the \textit{callback ReduceLROnPlateau} training parameter, with the intention of varying the learning rate throughout the training and overcoming the stagnation that could occur when using the same rate over many epochs. The best performing batch size was 128, and the model was subsequently reloaded and retrained in batch sizes of 96, 64, 32,16 and 8.

The following activation functions were tried: \textit{relu}, \textit{gelu}, \textit{selu}, \textit{mish}, \textit{swish} e \textit{lisht}. The \textit{mish} optimizer, created by \citep{DBLP:journals/corr/abs-1908-08681}, had a better performance than the others. Despite its high computational cost, its main advantage is that it avoids gradient saturation, which could cause an extreme slowdown in training, and also it has a strong regularization capacity, preventing overfitting.						

\section{Results}

\subsection{Interpretability}

\subsubsection{Intermediate Activation Maps}

The code provided by \citep{chollet2021deep} was used to visualize the intermediate activation maps of the convolutional and pooling layers for each class:

\begin{figure}[htb]
\makebox[\linewidth]{\includegraphics[width=1\linewidth]{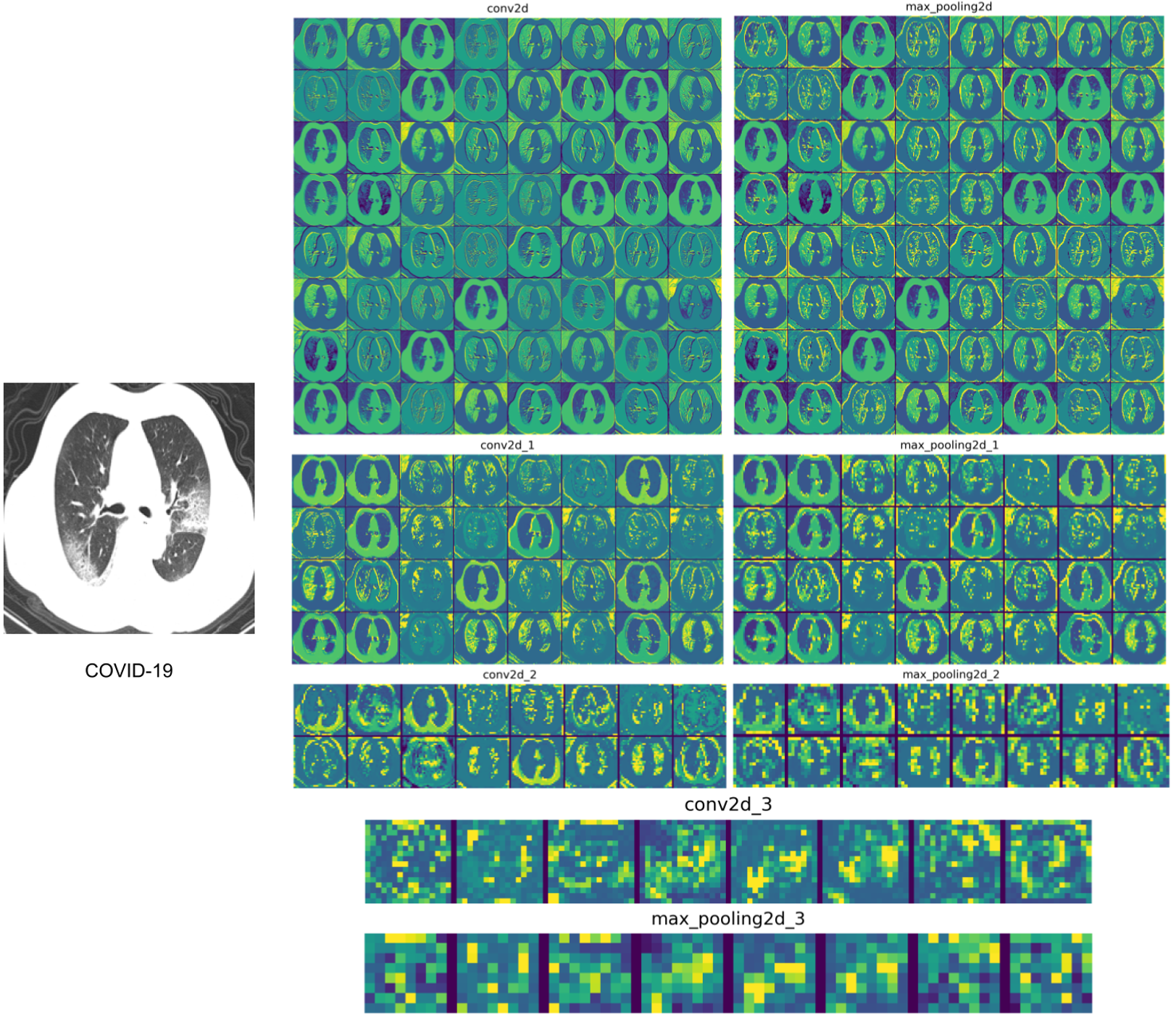}}
\caption{Activation maps for the COVID-19 class}
\end{figure}
\FloatBarrier

\begin{figure}[htb]
\makebox[\linewidth]{\includegraphics[width=1\linewidth]{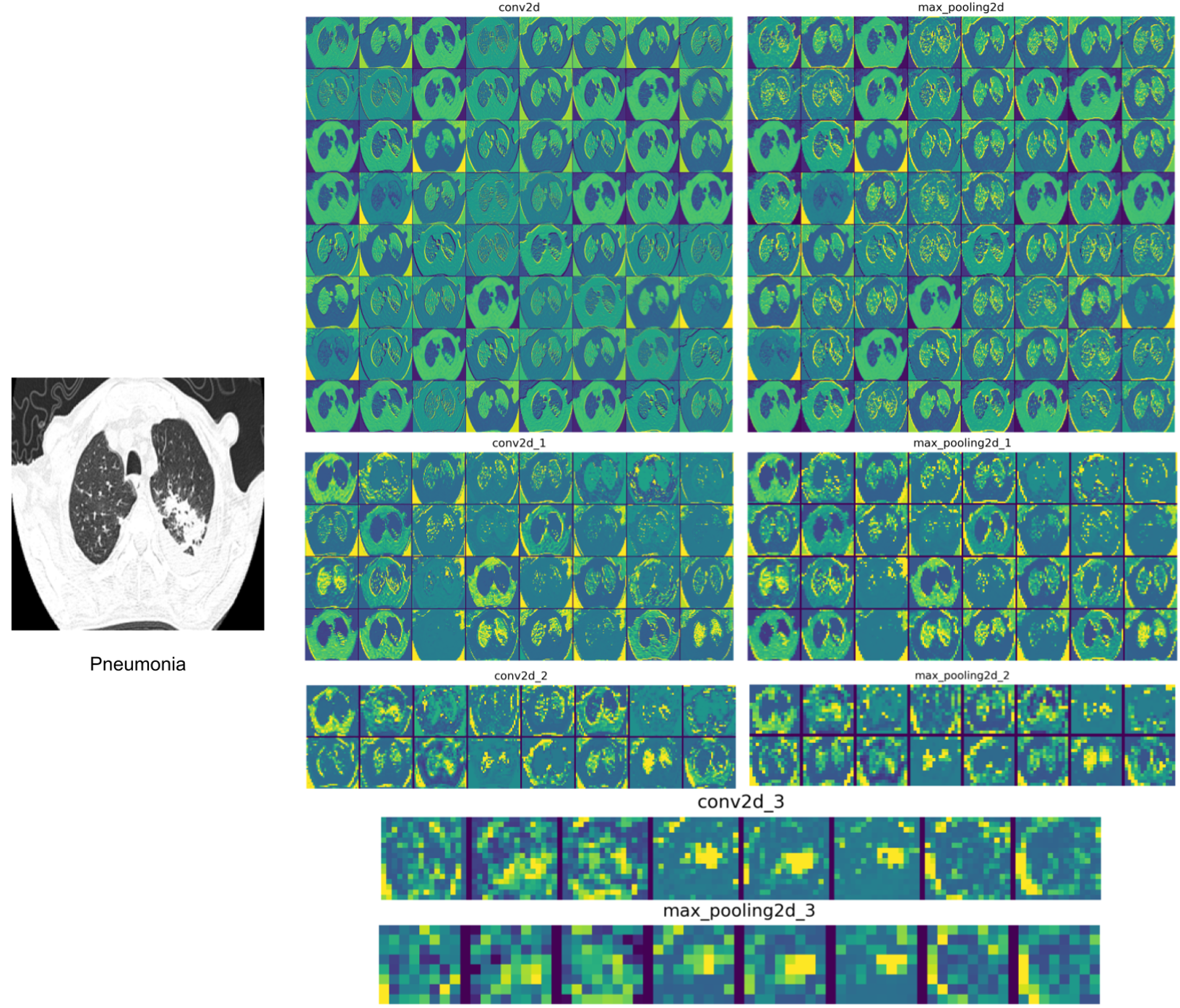}}
\caption{Activation maps for the Pneumonia class}
\end{figure}
\FloatBarrier

\begin{figure}[htb]
\makebox[\linewidth]{\includegraphics[width=1\linewidth]{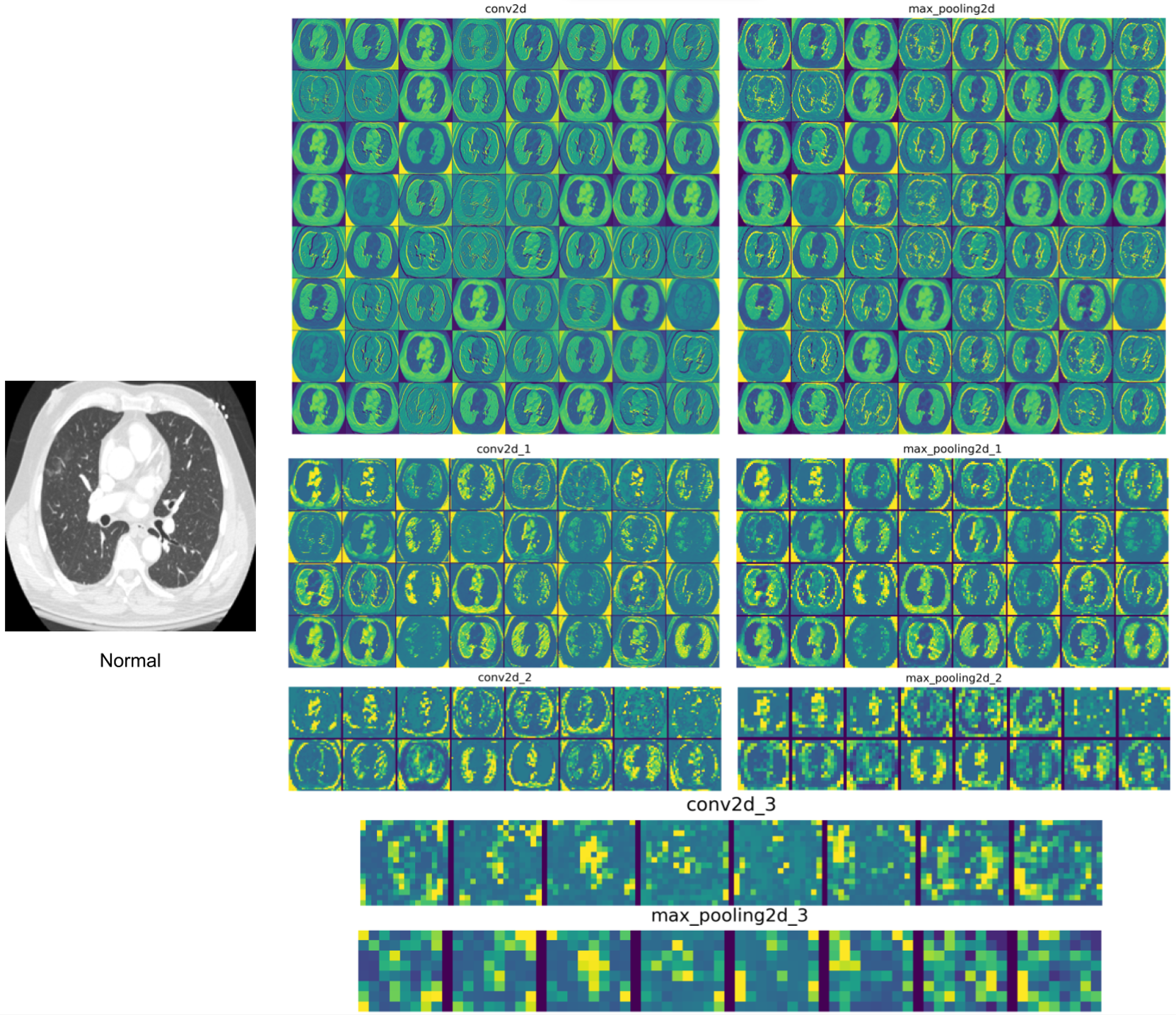}}
\caption{Activation maps for the Normal class}
\end{figure}
\FloatBarrier

\subsubsection{Gradient-weighted Class Activation Mapping (Grad-CAM)}

Figure 18 is based on \citep{keraschollet2021} and shows the Gradient-weighted Class Activation Mapping (Grad-CAM) technique applied to the images analyzed by the deep convolutional neural network. This technique uses a heatmap to show the points of greater importance for the neural network during the classification, the degree of importance follows the intensity of the heat map, red is the color used for the most important parts:

\begin{figure}[htb]
\makebox[\linewidth]{\includegraphics[width=0.8\linewidth]{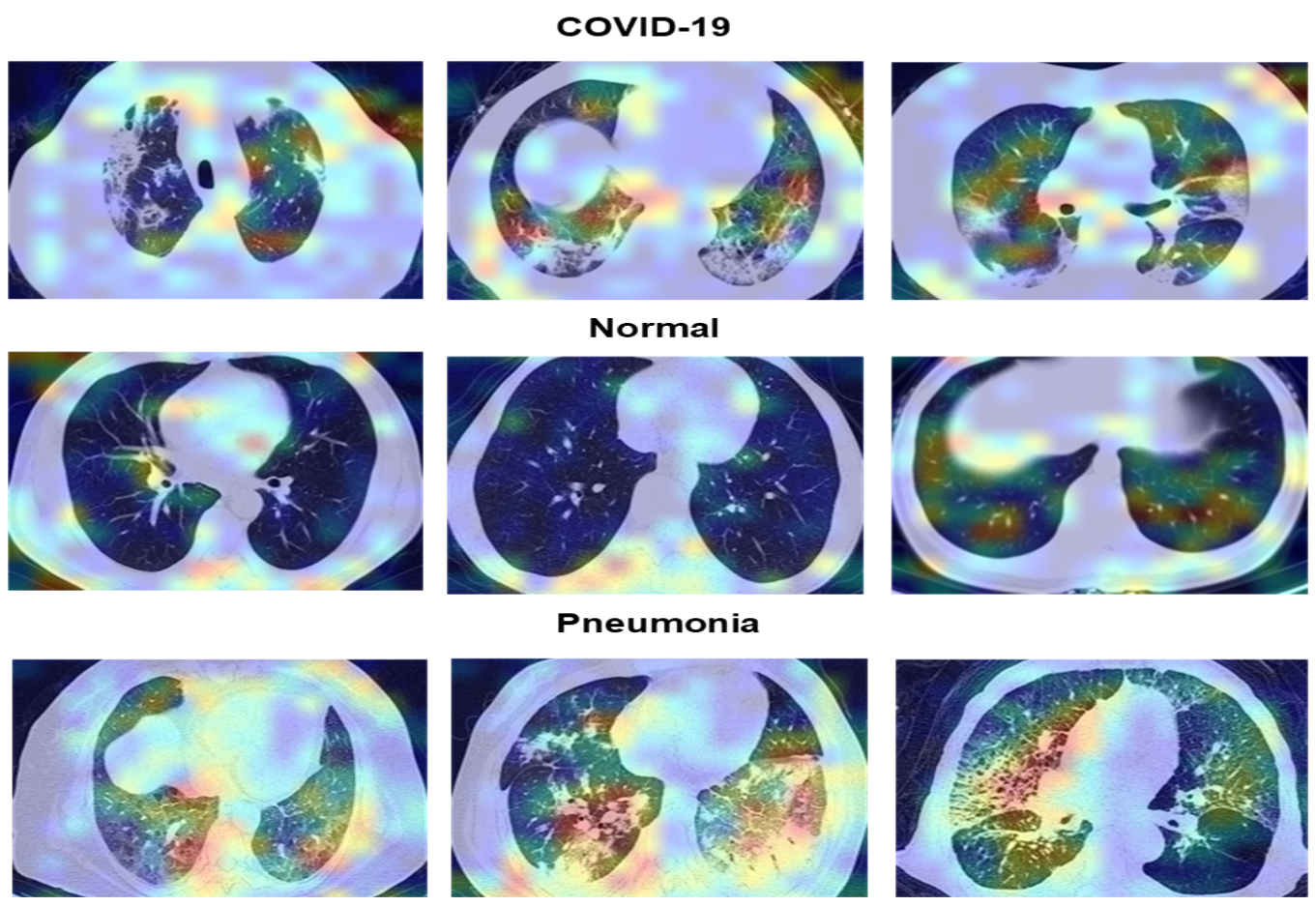}}
\caption{Grad-CAM of three examples of each class analyzed by the DCNNV-19 model}
\end{figure}
\FloatBarrier

\subsection{Performance}

The Confusion Matrix presented in Figure 19 indicates that the model was rarely confused in the classification of images of the Normal class, showing the highest error rate in the Pneumonia class, misclassifying 0.012 (or 1.2\%) of the images of this class as COVID-19:

\begin{figure}[htb]
\makebox[\linewidth]{\includegraphics[width=0.6\linewidth]{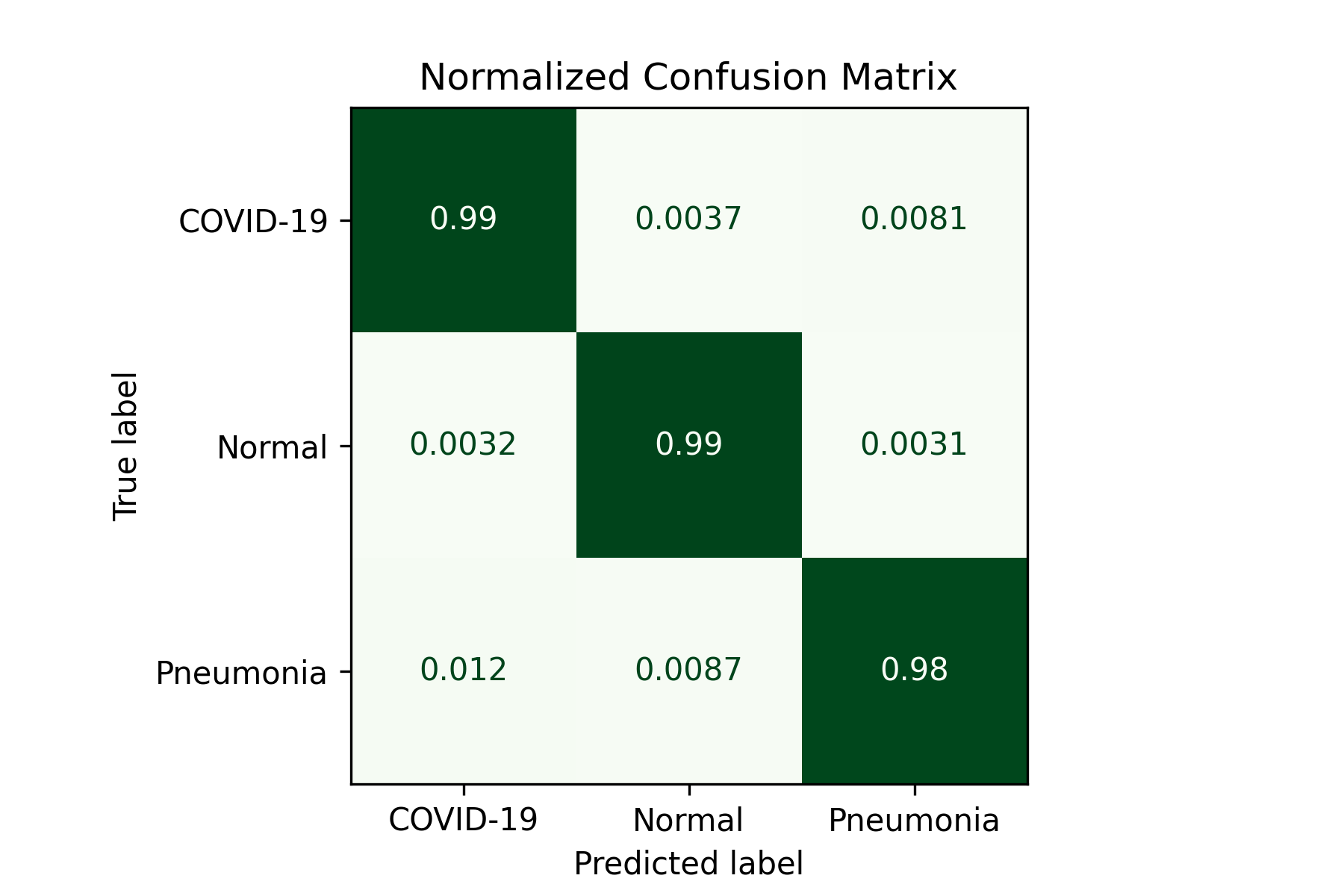}}
\caption{Normalized Confusion Matrix}
\end{figure}
\FloatBarrier

Since this is a multi-class problem, metrics such as accuracy, precision, or recall do not provide a score that shows the real performance of the model considering all classes. Therefore, Cohen's Kappa (Cohen's Kappa) coefficient was also used as an evaluation metric, which in this case measures the degree of agreement between the true labels and the predicted labels, and its score lies between -1 and 1. -1 represents complete disagreement among \textit{n} raters, 0 means agreement by chance, and a score of 1 represents complete agreement. On the test dataset, the model had an Accuracy of 0.9840, Loss of 0.0509, and Cohen's Kappa of 0.9759. Table 2 indicates the results of the model during all stages:

\begin{table}[htb]
\begin{center}
    \centering
    \begin{adjustbox}{width=0.5\textwidth}
    \begin{tabular}{ | c | c | c | c |}
    \hline
     & \textbf{Training} & \textbf{Validation} & \textbf{Test}\\ \hline
    \textbf{Loss} & 0.0128 & 0.0418 & 0.0509 \\ \hline
    \textbf{Accuracy} & 0.9958 & 0.9863 & 0.9840\\ \hline
    \textbf{Cohen's-Kappa} & 0.9937 & 0.9794 & 0.9759\\
    \hline
    \end{tabular}
    \end{adjustbox}
\end{center}
\caption{The score of the final model in the training, validation, and test stages}
\end{table}

When evaluating the model on the separate test dataset, Precision and Recall showed identical values for the same classes, generating a harmonic mean among these (F1-Score) with the same value. Macro and Weighted average were the same: 0.98 . Considering that this is a model for use in a medical task, and in this particular multi-class case, the most important metrics are the Recall of the COVID-19 and Pneumonia classes (0.98 in both), and the Accuracy of the Normal class (0.99), as indicated in Table 3. Whereas, assessing a patient with COVID-19 or Pneumonia as Normal would be a more serious situation than classifying a healthy person as sick.

\begin{table}[htb]
\begin{center}
    \begin{adjustbox}{width=0.5\textwidth}
   \begin{tabular}{ | c | c | c | c |}
    \hline
     & \textbf{COVID-19} & \textbf{Normal} & \textbf{Pneumonia}\\ \hline
    \textbf{Precision} & 0.98 & 0.99 & 0.98 \\ \hline
    \textbf{Recall} & 0.98 & 0.99 & 0.98\\ \hline
    \textbf{F1-Score} & 0.98 & 0.99 & 0.98\\ \hline
    \textbf{Support} & 7,395 & 7,395 & 7,395\\
    \hline
    \end{tabular}
    \end{adjustbox}
\end{center}
\caption{Final model score on Precision, Recall, and F1-Score}
\end{table}

\section{Conclusion}

With the increasing global number of cases of COVID-19, it has become necessary to search for new methods of diagnosis to provide rapid treatment so that the risk of hospitalization or serious health damage is minimized. Although RT-PCR has high sensitivity, its result is usually delivered in more than 24 hours to the patient, so it is not feasible in the severe symptoms cases. The created model achieved high accuracy on the test dataset according to several metrics: F1-Score of 98\%, 97.59\% Cohen's Kappa, 98.4\% Accuracy, and 5.09\% Loss. Thus, the proposed model demonstrates a capacity for use in real-world tasks, as a rapid and accurate analysis method in patients with symptoms of SARS or as a complementary method to RT-PCR when the latter is negative and SARS-CoV-2 infection is still suspected.

\newpage

\bibliographystyle{unsrtnat} 
\bibliography{references}

\end{document}